\documentclass[11pt]{article}
\usepackage[latin1]{inputenc}
\usepackage[english]{babel}
\usepackage[namelimits]{amsmath}
\usepackage{amssymb}
\usepackage{amsmath}
\usepackage{amsthm}

\begin{document}

\title{A quantum Friedmann flat spacetime: Uncertainty Relations, Thermodynamics and some cosmological consequences}
\author{
	Stefano Viaggiu,\\
	Dipartimento di Matematica,
	Universit\`a di Roma ``Tor Vergata'',\\
	Via della Ricerca Scientifica, 1, I-00133 Roma, Italy.\\
	E-mail: {\tt viaggiu@axp.mat.uniroma2.it}}
\date{\today}\maketitle

\begin{abstract}
We present Friedmann flat spacetime uncertainty relations (STUR) together with 
some cosmological implications. An interesting link between 
the Principle of
''gravitational stability against localization of events'' (PGSL) and the 
holographic Bekenstein entropy bound (HEB) is also
investigated. The same theorems leading to our STUR are used to calculate,
thanks to the holographic principle, 
the entropy of the universe at its apparent horizon. 
The generalized entropy formula can be used to discuss 
interesting links with a quantum spacetime.
\end{abstract}


\section{STUR for a Friedmann Flat Quantum Spacetime}

A non commutative spacetime 
has been derived in \cite{3} (DFR model) by a physically motivated principle. Namely, 
the DFR model \cite{3} is based on the 
''Principle of gravitational stability against localisation of events'' (PGSL):
\begin{itemize}
\item[] The gravitational field generated by the concentration of energy required
	by the Heisenberg uncertainty principle to localise an event
	in spacetime should not be so strong to hide the event itself to any
	distant observer - distant compared to the Planck scale.
\end{itemize}
In \cite{3}, by assuming the PGSL, the STUR in a minkowski spacetime has been obtained
with the use of the linear approximation of Einstein's equations and of the Heisenberg uncertainty principle. In \cite{4}
a stronger version than the one present in \cite{3} 
of the STUR in a minkowski spacetime, implying the DFR ones, has been derived and discussed.
In the cosmological case inequalities are at our disposal in the form of exact theorems that provide necessary or sufficient conditions that 
trapped surfaces do not form. 
A first class of results is present in \cite{6} while
in \cite{7} a necessary condition such that a generic equipotential spatial surface
$S$ with proper area $A$ be trapped can be found. Thanks to these theorems, 
it is reasonable to consider the following generalization of the Penrose's isoperimetric inequality for Friedmann flat expanding cosmologies: 
{\it black holes do not form if the (positive) excess of proper mass} $\delta M$ 
{\it inside a surface $S$ of proper area $A$ satisfies the inequality}:
\begin{equation}
\frac{\sqrt{A}}{4\sqrt{\pi}}+\frac{HA}{4\pi c} \geq \frac{G}{c^2}\delta M.\label{7}
\end{equation}
We can introduce \cite{8} a tetrad frame with spatial axes along the sides of the box.
An experimenter measures the proper time $t$ and the (infinitesimal) proper length at a given fixed time $t$ in terms of 
${\eta}^i= a(t) x^i$.
We can thus use the Heisenberg uncertainty relations in the proper coordinates $(t,{\eta}^i)$ and
the final form of our STUR becomes
\begin{eqnarray}
& &\frac{\sqrt{\Delta_\omega A}}{4\sqrt{3}}+\frac{\omega(H)\Delta_\omega A}{12c}\geq \frac{\lambda^2_P}{2} \frac{\Delta_\omega A}{\Delta_\omega V},\label{11}\\
& &c\Delta_\omega t\left(\frac{\sqrt{\Delta_\omega A}}{4\sqrt{3}}+\frac{\omega(H)\Delta_\omega A}{12c}\right)\geq \frac{\lambda^2_P}{2},\label{12}
\end{eqnarray}
where $\Delta_\omega V=\Delta_\omega \eta_1\Delta_\omega \eta_2\Delta_\omega \eta_3$ is the volume and $\Delta_\omega A$ 
the area of the localizing box (see \cite{8} for more details). 

\subsection{Physical consequences}
We study power law cosmologies with particle horizons, where  $a(t)=t^{\alpha},\;\alpha\in(0,1)$ and $H(t)=\alpha/t$. 
Classically, for an observer at (proper) time $t$ the particle horizon is the maximal proper spatial region that is in causal relation with him. 
This physical limit translated to a quantum level looks as follows:
\begin{equation}
\omega(H) \sqrt{\Delta_\omega A}\leq \sqrt{3}\frac{\alpha c}{(1-\alpha)}.\label{18}
\end{equation}
When the STUR are combined together with (\ref{18}), a maximum value for 
$H$ and $\rho$ arises:
\begin{equation}
\omega(H)\;\leq\frac{\alpha}{(1-\alpha)\sqrt{6(1-\alpha)}t_P},\;
\omega(\rho)_\textrm{max}=\frac{\alpha^2}{16\pi{(1-\alpha)}^3}{\rho}_P,\label{20}
\end{equation}
i.e. a strong indication that quantum mechanical effects prevent the appearance of a big bang singularity.
However, there exist another natural IR cutoff: the apparent horizon.

\section{Holographic entropy bound.} 
Bekenstein argument (see  \cite{8a} and references therein) suggests that
there exists an universal bound for the
entropy $S$ of a spherical object of radius $R$ and energy $E$ given by 
$S\leq S_{max}=\frac{2\pi k_B RE}{\hbar c}$. 
This bound has been refined by Susskind \cite{8b} (holographic entropy bound, HEB):\\
{\em The maximum entropy of a given region is provided by the biggest black hole fitting inside}. Note the analogy with the PGSL. Since $E\leq E_{max}=c^4 R/(2G)$ $\rightarrow$ 
$S_{max}= \frac{k_B A}{4 L_P^2}$. 
The situation is much more involved in a cosmological (dynamical) context. 
By means of the HEB,
we have proposed in \cite{8a,8ab} a generilized  Bekenstein-Hawking formula entropy for a black hole
embedded in a Friedmann flat universe. 
Thanks to the theorems in \cite{6,7},
we know that in an expanding spacetime
more energy with respect to the static case must be concentred to obtain a trapped surface. But more energy means more entropy.
By the same theorems used to obtain the STUR (\ref{11})-(\ref{12}), we have (see \cite{8a,8ab}):
\begin{equation} 
S_{BH}=\frac{k_B A}{4 L_P^2}+\frac{3k_B}{2c L_P^2}V H,
\label{2a}
\end{equation} 
where $V=4/3\pi L^3$ and $L$ the areal radius of the black hole. In the usual tractation, the term $\sim H$ is missing. 
The added term can be related to the
degrees of freedom caused by the expansion of the universe.
Since of the analogy between the PGSL (quantum spacetime) 
and the HEB, we expect some cosmological manifestation of non-commutative effects 
at the scale of the apparent horizon.  
 	
\section{Holographic principle and entropy of our universe}

What is the expression of the entropy of our universe? 
To this purpose, we can use the HEB.
In a cosmological context event horizons become teleological objects that are attainable only for eternal observers.
In \cite{9,10} it has been proposed that the relevant concept identifying black holes in a non-stationary context is provided by 
apparent horizons. An apparent horizon is defined to be
a marginally trapped surface with vanishing expansion $\theta$ for outgoing light rays. 
All Friedmann spacetimes with positive energy density have 
an apparent horizon at a given proper areal radius $L_h$: in the flat case $L_h=c/H$.
The apparent horizon is the thermodynamic radius of the universe. 
It is thus natural to suppose that the HEB is exactly saturated at the apparent horizon $L_h$
\begin{equation} 
S_h=\frac{k_B A_h}{4 L_P^2}+\frac{3k_B}{2c L_P^2}V_h H,
\label{23}
\end{equation}
where $V_h=4/3\pi L_h^3$.
The formula (\ref{23}) was generilized to non-flat Friedmann universes in \cite{8ab}.

\section{Thermodynamic at the apparent horizon}
Since the universe is thermodynamically defined at $L=L_h$, we can write down the first law
$T_h dS_h=dU_h+P_h dV_h$.
To an apparent horizon can be associated a temperature $T_h$ that is (opportunely normalized)
$T_{h}=\frac{\hbar H}{4\pi k_B}$.
After differentiating the (\ref{23}) we have:
\begin{equation}
dU_h=\frac{c^4}{2G}dL_h+\frac{c^3}{2G}L_h^2dH,\;\;
P_h=\frac{3 c^3 H}{8\pi G L_h}.\label{26}
\end{equation}
\begin{itemize}
\item When the (\ref{26}) is applied at $L_h$, we have $dU_h=0$, i.e. the internal energy of our universe is exactly conserved at
$L_h$. From dimensional arguments, with only the constant $G,c,\Lambda$ present in the Hilbert-Einstein action,
since the energy of minkowski
spacetime ($\Lambda=0$) is zero, we conclude that $U_h=0$. 
The internal energy of our universe is zero! This is an hold conjecture \cite{12}, but a rigorous proof is still lacking. 
\item In our context, the Helmholtz energy $F$ is $F_h\sim H_{,t}/H^2$ and is stationary only
for a de Sitter universe. 
Hence a de Sitter universe is in thermal equilibrium with its holographic screen with
$L_h=const$ and thus $T_u=T_h$, where $T_u$ is the temperature of the universe.
\end{itemize}

\section{Some cosmological implications}
We analyze some interesting cosmological implications of the results above in terms of a quantum spacetime.
\begin{itemize}
\item The internal energy $U$ is the energy necessary to create a given system. 
We have seen that our STUR imply a maximum allowed value for $\omega(H)$. What
generated a Friedmann spacetime?
Since our spatially flat 
universe has effectively zero internal energy, our universe could be emerged from nothing. 
More precisely, a Friedmann flat quantum spacetime could be emerged from a minkowski quantum spacetime. 
An initial cosmological constant ${\Lambda}_i$ can be emerged thanks to the repulsive effect of a 
minkowski quantum spacetime at the Planck length ${\lambda}_P$ \cite{13}
\footnote{Since the apparent horizon is spherical it is resonable to consider
localizing state with all the $\Delta x^{\mu}$ of the same order.}. 
After introducing the Planck constant $\hbar$, the energy displacement
$U_h$ of a Friedmann universe can be $U_h=kc\hbar\sqrt{\Lambda}_i$, with $k$ a positive constant. 
A primordial 
quantum Friedmann flat spacetime can thus be emerged from a minkowski one if an uncertainty (accuracy) $kc\hbar\sqrt{\Lambda}_i$
of energy 
is allowed in minkowski spacetime in a given measurement during a time $\Delta t$.
From quantum mechanics, we know that
an uncertainty of a quantity $kc\hbar\sqrt{\Lambda}_i$ in the energy is possible iff:
\begin{equation}
{\Delta} t kc\sqrt{\Lambda}_i\geq \frac{1}{2}.               
\label{27}                
\end{equation}
By taking a 'spherical' localizing state ${\omega}_s$ (see \cite{3}) in a minkowski spacetime saturing the inequality (\ref{27}) with 
$c\Delta t\sim \Delta x\sim {\lambda}_P$, we obtain a maximum ${\Lambda}_{imax}$ given by
${\Lambda}_{imax}\sim \frac{1}{4k^2{\lambda}_P^2}$. Note that the maximum for
 $\omega(H)$ implies that ${\Lambda}_{imax}=1/{\lambda}_P^2$, i.e. 
$k=1/2$. As expected, a huge cosmological constant can emerge
thanks to non commutative quantum effects. 
Note that with the same argument but applied to the current cosmological time
$t_0$ in (\ref{27}) and taking the infrared (IR) natural cutoff 
given by the the actual apparent horizon $c/H_0\sim t_0$, we obtain ${\lambda}_{0min}=1/(4k^2 c^2t_0^2)$, and by taking $k=1/2$ (or of the order of unity) we obtain
a value of the cosmological constant practically coincident with the actual one (holographic dark energy). This is a 
well known fact, but in our context
assumes a new interesting interpretation.
\item As stated above, a de Sitter phase emerges when $T_u=T_h$. Thanks to the maximum allowed value for $\omega(H)$, the condition 
$T_u>T_h$ is always satisfied in our universe, and usual thermodynamic laws
are fulfilled for a universe filled with ordinary matter-energy fluids.
\item After the emergence of a Friedmann flat universe,
we may think to a universe filled with micro black holes where Hawking evaporation stops at the Planck length \cite{13}. 
The temperature of the universe
decreased more quickly with respect to $T_h$ and a de Sitter phase emerges when $T_u=T_h$. By means of (\ref{23})
(see \cite{8a}), we can evaluate the behaviour of $H$ soon after the beginning of the inflationary era;
\begin{equation}
H=H_I-\frac{c^2{\lambda}_P^2}{240\pi L_I^4}\left(t-t_I\right)+o(1),
\label{28}
\end{equation}
where $t\in[10^{-34},10^{-32}] s$ ($t_F$ denotes the end of the inflation). We have, according to the Planck data, 
a red-shifted inflation. The ratio ${\lambda}_P^2/L_I^4$ measures the non gaussianity level of the perturbations.

\end{itemize}


\begin{thebibliography}{0}
\bibitem{3} S. Doplicher, K. Fredenhagen and J. E. Roberts, {\em Comm. Math. Phys.} {\bf 172}, 187 (1995).
\bibitem{4}L. Tomassini and S. Viaggiu, {\em Class. Quantum Grav.} {\bf 28}, 075001 (2011).
\bibitem{6} U. Brauer, E. Malec and N.O. Murchadha, {\em Phys. Rev.}{D\bf  49}, 5601 (1994).
\bibitem{7}P Koc and E Malec E, {\em Acta Phys. Pol.} {B\bf 23}, 123 (1992).
\bibitem{8}L. Tomassini and S. Viaggiu, {\em Class. Quantum Grav.} {\bf 31}, 185001 (2014). 
\bibitem{8a}S. Viaggiu, {\em Mod. Phys. Letters} {A\bf 29}, 1450091 (2014).
\bibitem{8b}L. Susskind, {\em J. Math. Phys.}{\bf 36}, 6377 (1995).
\bibitem{8ab}S. Viaggiu, {\em Gen. Relativ. Gravit.}, issue 8 {\bf 47} (2015) (arXiv:1506.08573).
\bibitem{9} S. A. Hayward, {\em Phys. Rev.} {D\bf 49}, 6467 (1994).  
\bibitem{10}S. A.  Hayward,  {\em Class. Quantum Grav.} {\bf 11}, 3025 (1994).  
\bibitem{12}E. P. Tryon, {\em Nature} {\bf 246}, 396 (1973). 
\bibitem{13}D. Bahns, S. Doplicher, G. Morsella and G. Piacitelli, 
{\em Advances in Algebraic Quantum Field Theory},
289-330 Springer (2015).
(arXiv:1501.03298).
\end{thebibliography}
\end{document}